\documentstyle[aps,multicol,epsf]{revtex}

\begin{document}
\draft

\title{
WWW and Internet models from 1955 till our days  
and  \\
the ``popularity is attractive'' principle
}

\author{
S.N. Dorogovtsev$^{1, 2, \ast
}$, J.F.F. Mendes$^{1,\dagger}$, 
and A.N. Samukhin$^{2, \ddagger}$
}

\address{
$^{1}$ Departamento de F\'\i sica and Centro de F\'\i sica do Porto, Faculdade 
de Ci\^encias, 
Universidade do Porto\\
Rua do Campo Alegre 687, 4169-007 Porto, Portugal\\
$^{2}$ A.F. Ioffe Physico-Technical Institute, 194021 St. Petersburg, Russia 
}

\maketitle

\begin{abstract}
We note that the model discussed in the communication of S. Bornholdt and H. Ebel 
(World Wide Web scaling exponent from Simon's 1955 model, cond-mat/0008465) 
is the particular case of the model considered and solved exactly in our paper, 
cond-mat/0004434. These models may be used for estimation of the order of the deviation 
of the scaling exponent from $2$ both for the distributions of incoming links and links coming out from nodes but not for the obtaining some specific values of the exponents from the WWW growth data. 
We emphasize that, unlike the statement of Bornholdt and Ebel, 
both the network under consideration and the model of Barab\'{a}si and Albert provide quite equal possibilities for individual growth. 
There is no great difference between them in this respect.  
The resulting distributions for individual nodes and arising scaling relations have been obtained in our paper, cond-mat/0004434. 
We discuss briefly the modern state of art in the 
physics 
of the evolving networks
and the great role of the general principle -- {\em popularity is attractive} -- in the self-organization of complex communications networks, in physics of nonequilibrium phenomena, and in Nature.
\end{abstract}

\pacs{
}

\begin{multicols}{2}

\narrowtext


An interesting linking between the Simon's model \cite{s55} (1955) and new conceptions of the developing networks was made in the communication of Bornholdt and Ebel \cite{be00}. Using a simple model of the growing network with preferential linking and referring to the Simon's model Bornholdt and Ebel have obtained the connectivity distribution $P(k) \propto k^{-\gamma}$ with the exponent $\gamma$ which value have coincided with the experimental result (2.1) \cite{ajb99,bkm00} for the exponent of the distribution of incoming links in World-Wide Web.  

Here, we note that this model is the particular case of the model which was considered and solved exactly (for large networks) in our paper \cite{dms00}. The only difference is the different definitions of the time scales that do not influence the result. We show that these models may explain the scales of the deviation of the exponent $\gamma$ from $2$ for both the distribution of incoming links in the World Wide Web and the distribution of links coming out from nodes but they can not provide the particular experimental values of these exponents. We demonstrate these estimations and explain their quality.

We point out that the statement of Bornholdt and Ebel that this model ``allows for different growth rates'' for nodes with same connectivity and, e.g., the model of Barab\'{a}si and Albert \cite{ba99} do not allow is not right. 
Both models provide quite equal possibilities in this respect. 
Their connectivity distributions of individual nodes are of a similar type. 

We describe the connections of the idea of the preferential linking with classical problems of statistical physics of nonequilibrium phenomena and discuss briefly the state of art in the statistical physics of evolving networks. 
 
Also, the aims of the present comment are to explain popularly the results of our papers \cite{dms00,dm00,dm001} obviously written in a too mathematical language to be noticed and to manifest quite obvious simple ideas. 

Below we:

\noindent
(i) describe the model and its particular cases, 

\noindent
(ii) explain what estimates for the World Wide Web exponents it may give, 

\noindent
(iii) explain what kind of the distributions of connectivity of individual sites does the preferential linking produce, 

\noindent
(iv) discuss the modern state of art in understanding of the World Wide Web, Internet, and other evolving networks, 
and, finally,

\noindent
(v) demonstrate the simplest derivation of the expression for the exponent of the connectivity distribution, Eq. (\ref{a1}).     

\vspace{5pt}
{\em I. The model}.---At each increment of time a new node is created. Simultaneously $m$ {\em directed} links is distributed among all the nodes of the network by the following rules: 

(i) The source of any of these links may be anywhere; 

(ii) The target ends of these links are attached to nodes chosen preferentially: 
Probability to chose some particular node is proportional to $k+A$. 
Here, $k$ is the number of the distributed preferentially links which 
point to this node, and $A$ is some positive constant. 

The distribution of incoming links \cite{note} in this model is of the power-law form with the exponent $\gamma=2+A/m$ \cite{dms00}. We call $A$ the {\em initial attractiveness} of nodes (for links). $A$ is a parameter of the problem and is any positive constant. 
In particular, if $n$ links is already directed to any newborn node, $A=n+B$, where $B$ is another constant. 

We stress that the preferential linking with the probability proportional to $k+const$ 
is the {\em general} type of the preferential linking producing free-scale networks (i.e., the networks with power-law connectivity distributions). E.g., the rule with the linking probability proportional to $k^a, \ a\neq 1$, do not give scale-free networks \cite{krl00}. Hence, we consider the most general situation.     

In addition to links distributed preferentially, at each increment of time, one may also distribute some links, $n_r$, randomly without any preference (always there exist some crazy guys which 
make their references to God knows what). 
Hence, \\
--- $m$ is the number of links distributed preferentially each unit of ``time'',\\ 
--- $n$ is the number of links pointed at a new node at the instant of its birth, $n \geq 0$, \\
--- $n_r$ is the number of links distributed without preference at each increment of time, $n_r \geq 0$, \\
--- $B$ is any constant such that $n+B>0$, \\
--- one node is born per unit of ``time''. 
 
With these definitions, the expression for the exponent is 

\begin{equation}
\gamma = 2 + \frac{n_r+n+B}{m}
\, .   
\label{a1}
\end{equation} 
Note that all the factors accounted for in Eq. (\ref{a1}) (see the numerator), in fact, play the same role.  
An exact form of the connectivity distribution was obtained in \cite{dms00}. Below we show how 
our relation, Eq. (\ref{a1}), may be obtained in a most simple way using continuous approach. We have demonstrated that this approach gives exact results for the scaling exponents \cite{dms00,dm00,dm001}. 

Bornholdt and Ebel considered the model in which, at each increment of time, one link is added to a network. 
In our case, time is equal to the total number of nodes.
If we rescale time and set $n_r=0$, $n=1$, and $B=0$, we get the particular case considered by Bornholdt and Ebel with their probability $\alpha=1/(1+m)$ (one may check that all the quantities may be taken noninteger). If we set $n_r=0$ and $n+B=m$, we get, in fact, the Barab\'{a}si-Albert's model.  

\vspace{5pt}
{\em II. Possible estimations for the World Wide Web}.---We do not know values of any of the quantities from Eq. (\ref{a1}). The constant $B$ may takes {\em any} values, the number of the randomly distributed links, $n_r$, may be not small (many crazy guys), $n$ is not fixed. From the experimental data \cite{bkm00}, we know more or less the sum $m+n+n_r \sim 10 \gg 1$ (between 7 and 10, more precisely), and that is all.

The only thing we can do, it is to fix the scales of the quantities. The natural characteristic values for $n_r+n+B$ in Eq. (\ref{a1}) are (a) $0$, (b) $1$, (c) $m \gg 1$, and (d) infinity. In the first case, node have zero initial attractiveness, and all new links are directed to the oldest node, $\gamma \to 2$. In the last case, there is no preferential linking, and the network is not scale-free, $\gamma \to \infty$. Let us consider the really important cases (b) and (c).

{\bf (b)} \ How do pages appear in the Web? Suppose, you want create your personal home page. Of course, first you prepare it, put references, etc. But that is only the first step. You have to make it accessiable in the Web, to lanch it. You come to your system administrator, he put a reference to it (one reference) in the page of your institution, and that is more or less all -- your page is in the World Wide Web. If the process of appearing of each document in the Web is as simple as the creating of your page -- only one reference to the new document $(n=1)$ -- and if one forgets about the terms $n_r$ and $B$ in  Eq. (\ref{a1}), 
than, for the exponent of the distribution of the incoming links, we would get immediately the same estimation as in \cite{be00}, $\gamma-2 \sim 1/m \sim 10^{-1}$. 
This estimation is indeed coincides with experimental value $\gamma_{in}-2=0.1$ \cite{bkm00}.
We repeat again, that 
it follows only from the fixation of scale of the involved quantities. 
We emphasize that there are no any general reasons to set, e.g., $B=0$. 
A lot of real processes are not included in this estimation. Aging of nodes changes $\gamma$ \cite{dm00,asbs00}, account for dying of nodes $\gamma$ \cite{dm001} (the half-life of a page in the Web is of the order of half a year) changes $\gamma$. The ratio between the total number of links and the number of nodes in the Web is not constant \cite{bkm00}, it increases with time, the growth of the Web is nonlinear. This factor also change the value of $\gamma$, in future it may become even lower than $2$ \cite{dm002}.  

{\bf (c)} \ Above we discussed the distribution of incoming links. Eq. (\ref{a1}) may be also applied for the distribution of links which come out from documents of the Web. 
In this case all the quantities in Eq. (\ref{a1}) takes other values which are again unknown. Nevertheless, one may think that the number of the links distributing without any preference, $n_r$, is not small now. Even beginners proceed with linking of their pages. 
Hence, $n_r \sim m$ -- we have no another available scale, -- and $\gamma-2 \sim m/m \sim 1$. 
We again compare this estimation with the experimental value, $\gamma_{out}-2=0.7$ \cite{bkm00} and find that we have made only small mistake. 
Of course, this is not quite fair but now we see what estimations may be made. 

What may be done to improve these estimates? As we saw, any experimental information about the correlation between the numbers of in- and out- links of nodes would be useful.    

\vspace{5pt}
{\em III. Scaling of the distributions of connectivity of individual nodes}.---
One may study not only the global connectivity distribution of the network but the connectivity distributions of individual nodes. We have obtained the exact expressions for these distributions of the model under discussion \cite{dms00}. Unlike the statement of Bornholdt and Ebel that this model provides some extra possibilities for different growth rate, we found that these distributions are of the same type as 
in the model of Barab\'{a}si and Albert \cite{ba99}. 
It is a mistake to think that the network under consideration may demonstrate the growth crucially different to the Barab\'{a}si-Albert's model. 
Indeed, the last one is only the particular case of the introduced model but this is not a marginal case to produce a radical difference. 
Therefore, the ``rich-get-richer'' behavior also present in 
the model under consideration, although all these networks provide 
broad distribution of the following scaling form:   

\begin{equation}
p(k,s,t) = \left(\frac{s}{t}\right)^\beta f\left(k\left(\frac{s}{t}\right)^\beta\right)
\, .  
\label{a2}
\end{equation} 
Here, $p(k,s,t)$ is the connectivity distribution of the node appeared at time $s$ measured at time $t$ (i.e., when the size of the network is $t$), $s<t$. The scaling exponent $\beta$ is related to $\gamma$ by the following scaling relation:

\begin{equation}
\beta(\gamma-1) = 1
\, ,  
\label{a3}
\end{equation}
that was obtained in \cite{dm00}. It was proved in \cite{dms00}, that this relation is general and is valid for all such scale-free networks. We found that the scaling function is broad, $f(x)=x^{A-1}\exp(-x)$, so the connectivity distributions of individual nodes are broad. They are narrow only in the continuous approximation. 

\vspace{5pt}
{\em IV. The principles of communications networks}.---Perhaps, 
the greatest results in the theory of communications networks belong 
to ``the farther'' of the Internet, P. Baran, (1964) \cite{b64}, so they were obtained long ago. Nevertheless, the idea of the preferential linking in application to the evolving networks is really strong. Indeed, one of the first questions concerning the communications networks -- {\em why are they scale-free?} 

Two answers have been given quite recently. First, -- because of the preferential linking \cite{ba99}. The growing networks are self-organized into the scale-free structures because {\em popularity is attractive.} This general principle governs the process of the growth of the World Wide Web! 

Second, --- because otherwise they would be unstable, weak against processes of decay, and could not exist as united systems \cite{ba00a}. It has been shown that only the scale-free networks with the exponent $\gamma$ less than $3$ are resilient to random breakdowns \cite{ceah00}. 
The ``infinite'' scale-free networks with $\gamma < 3$ do not decay for {\em any} concentration (less than one) of randomly removed nodes or links. 

Therefore, the evolution of the World Wide Web is governed by one of the most general principles in Nature.
Only lazy can not imagine its numerous realizations 
in natural and social sciences, and in economics. 
It certainly establishes not only networks of movie actors. 
In particular, in physics, it produces various processes of growth and fragmentation. 
For instance, in \cite{dms00} we studied the growth of the networks mathematically as the problem of the preferential distribution of new particles among growing number of places. 

Now the World Wide Web is the most striking product of the principle {\em popularity is attractive} but it is only one of its products.      
It is wonderful that even in 1955 people thought about the quantitative formulation of this global principle.

\vspace{5pt}
{\em V. Derivation of our expression for the exponent}.---Let us show in the simplest way how one may obtain Eq. (\ref{a1}). We use the continuous approximation which gives exact results for the scaling exponents \cite{dms00}. Using this approach, one may write the following equation with the boundary condition for the average connectivity (here, average number of incoming links) of the node $s$ at time $t$:

\begin{eqnarray}
\frac{\partial\overline{k}(s,t)}{\partial t}  & = &
\frac{n_r}{t}  +   
m\,\frac{\overline{k}(s,t)+B}{\int_0^t du[\overline{k}(u,t)+B]}   
 , 
\nonumber
\\
[5pt]  
\overline{k}(t,t) & = &  n
\, . 
\label{1}
\end{eqnarray}  
The first term in the right hand part of it arises from the links distributed without preference, the second one -- from the preferential linking. 
Integrating Eq. (\ref{1}) from $0$ to $t$ we get

\begin{equation}
\int_0^t ds\, \frac{\partial \overline{k}(s,t)}{\partial t} =
\frac{\partial}{\partial t} \int_0^t ds\,\overline{k}(s,t) - \overline{k}(t,t) = 
n_r+m
\,   
\label{2}
\end{equation} 
and the obvious relation,

\begin{equation}
\int_0^t ds\, \overline{k}(s,t) = 
(n_r+m+n)t
\, .  
\label{3}
\end{equation} 
Substituting Eq. (\ref{3}) into Eq. (\ref{1}) we obtain the equation

\begin{equation}
\frac{\partial \overline{k}(s,t)}{\partial t} = 
\frac{n_r}{t}  + \frac{m}{m+n_r+n+B}\,\frac{\overline{k}(s,t)+B}{t}
\,   
\label{4}
\end{equation} 
the solution of which in the scaling region is of the following form: 

\begin{equation}
\overline{k}(s,t) \propto \left(\frac{s}{t}\right)^{-\beta}
\,   
\label{5}
\end{equation}
(we used the boundary condition $\overline{k}(t,t) = n$) with the scaling exponent

\begin{equation}
\beta=\frac{m}{m+n_r+n+B}
\, .  
\label{5p}
\end{equation} 
Finally, using the scaling relation, Eq. (\ref{a3}), we obtain 
$\gamma = 1+1/\beta = 2 + (n_r+n+B)/m$. 
\\


\noindent
$^{\ast}$      E-mail address: sdorogov@fc.up.pt\\
$^{\dagger}$   E-mail address: jfmendes@fc.up.pt\\
$^{\ddagger}$  E-mail address: alnis@samaln.ioffe.rssi.ru


\end{multicols}


\begin{references}
\bibitem{note} Note that the power law is the same whether we calculate it for 
the distribution of all incoming links of nodes or only for the distributed preferentially part of them.
\end{references}
\end{document}